\documentclass[aip,superscriptaddress,reprint]{revtex4-1}

\usepackage[T1]{fontenc}       
\usepackage{textgreek}
\usepackage{ifthen}
\usepackage{amstext}
\usepackage{amssymb}
\usepackage{graphicx}
\usepackage{color}
\usepackage{soul}

\newcommand*\m[1]{\mathrm{#1}}

\let\v\undefined

\newcommand*\v[1]{\vec{#1}}

\newcommand*\f[2]{\frac{#1}{#2}}
\newcommand*\beq{\begin{equation}}
\newcommand*\eeq{\end{equation}}
\newcommand*{\dfp}[3][]{\ifthenelse{\equal{#1}{}}{\frac{\partial #2}{\partial #3}}{\left.\frac{\partial #2}{\partial #3}\right|_{#1}}}

\newcommand*{\lr}[2][]{\ifthenelse{\equal{#1}{}}{\left(#2\right)}{\ifthenelse{\equal{#1}{[}}{\left[#2\right]}{\ifthenelse{{\equal{#1}{\{}}\or{\equal{#1}{\}}}}{\left\{#2\right\}}{\ifthenelse{{\equal{#1}{(}}\or{\equal{#1}{)}}}{\left(#2\right)}{\ifthenelse{{\equal{#1}{\langle}}\or{\equal{#1}{\rangle}}}{\left\langle#2\right\rangle}{\ifthenelse{{\equal{#1}{|}}}{\left\lvert#2\right\rvert}{#2}}}}}}}

\begin{document}

\title{Combined electrical transport and capacitance spectroscopy of a $\m{MoS_2-LiNbO_3}$ field effect transistor}

\author{Wladislaw Michailow}
\email{wladislaw.michailow@gmx.de}
\affiliation{Lehrstuhl f\"{u}r Experimentalphysik 1 and Augsburg Centre for Innovative Technologies (ACIT), Universit\"{a}t Augsburg, Universit\"{a}tsstr. 1, 86159 Augsburg, Germany} 

\author{Florian J. R. Sch\"ulein}
\affiliation{Lehrstuhl f\"{u}r Experimentalphysik 1 and Augsburg Centre for Innovative Technologies (ACIT), Universit\"{a}t Augsburg, Universit\"{a}tsstr. 1, 86159 Augsburg, Germany} 
\affiliation{Nanosystems Initiative Munich (NIM), Schellingstr. 4, 80799 M\"{u}nchen, Germany}

\author{Benjamin M\"oller}
\affiliation{Lehrstuhl f\"{u}r Experimentalphysik 1 and Augsburg Centre for Innovative Technologies (ACIT), Universit\"{a}t Augsburg, Universit\"{a}tsstr. 1, 86159 Augsburg, Germany} 

\author{Edwin Preciado}
\affiliation{Chemistry, Materials Science \& Engineering and Electrical Engineering, University of California, Riverside, California 92521, United States}

\author{Ariana E. Nguyen}
\affiliation{Chemistry, Materials Science \& Engineering and Electrical Engineering, University of California, Riverside, California 92521, United States}

\author{Gretel Von Son}
\affiliation{Chemistry, Materials Science \& Engineering and Electrical Engineering, University of California, Riverside, California 92521, United States}

\author{John Mann}
\affiliation{Department of Physics, Pepperdine University, Malibu, California 90263, United States}

\author{Andreas L. H\"orner}
\affiliation{Lehrstuhl f\"{u}r Experimentalphysik 1 and Augsburg Centre for Innovative Technologies (ACIT), Universit\"{a}t Augsburg, Universit\"{a}tsstr. 1, 86159 Augsburg, Germany} 

\author{Achim Wixforth}
\affiliation{Lehrstuhl f\"{u}r Experimentalphysik 1 and Augsburg Centre for Innovative Technologies (ACIT), Universit\"{a}t Augsburg, Universit\"{a}tsstr. 1, 86159 Augsburg, Germany} 
\affiliation{Nanosystems Initiative Munich (NIM), Schellingstr. 4, 80799 M\"{u}nchen, Germany}

\author{Ludwig Bartels}
\affiliation{Chemistry, Materials Science \& Engineering and Electrical Engineering, University of California, Riverside, California 92521, United States}

\author{Hubert J. Krenner}
\email{hubert.krenner@physik.uni-augsburg.de}
\affiliation{Lehrstuhl f\"{u}r Experimentalphysik 1 and Augsburg Centre for Innovative Technologies (ACIT), Universit\"{a}t Augsburg, Universit\"{a}tsstr. 1, 86159 Augsburg, Germany} 
\affiliation{Nanosystems Initiative Munich (NIM), Schellingstr. 4, 80799 M\"{u}nchen, Germany}

\begin{abstract}
We have measured both the current-voltage ($I_\m{SD}$-$V_\m{GS}$) and capacitance-voltage ($C$-$V_\m{GS}$) characteristics of a $\m{MoS_2-LiNbO_3}$ field effect transistor. From the measured capacitance we calculate the electron surface density and show that its gate voltage dependence follows the theoretical prediction resulting from the two-dimensional free electron model. This model allows us to fit the measured $I_\m{SD}$-$V_\m{GS}$ characteristics over the \emph{entire range} of $V_\m{GS}$. Combining this experimental result with the measured current-voltage characteristics, we determine the field effect mobility as a function of gate voltage. We show that for our device this improved combined approach yields significantly smaller values (more than a factor of 4) of the electron mobility than the conventional analysis of the current-voltage characteristics only. 
\end{abstract}

\maketitle

After the rise of graphene\cite{Novoselov04, Novoselov05, Zhang05}, a wide range of two-dimensional (2D) materials\cite{Novoselov05-2D} shifted into focus of fundamental and applied research\cite{roadmap-graphene}.
One particularly important class of 2D materials are transition metal dichalcogenides (TMDs)\cite{overview-article}.
One important representative TMD is molybdenum disulfide, $\m{MoS_2}$, whose indirect band gap changes to a direct one when its thickness is reduced to one single monolayer \cite{mos2-bandgap2, Splendiani2010}. The resulting high optical activity and sizable bandgap of $\sim\,1.9\,\m{eV}$ make this material ideally suited for optoelectronic applications\cite{Lopez-Sanchez2013} and, thus the optical and electronic properties of $\m{MoS_2}$ and related materials have been investigated intensively in the last years\cite{Sun2016}.
In particular, field effect transistors (FETs) and logical circuit prototypes have been devised and realized\cite{Radisavljevic2011,Radisavljevic2011Integrated,Fiori2014}. 
In such devices, source and drain contacts are patterned onto the TMD film, and the charge carrier density is controlled by gate contacts. For FET devices, the transport mobility of the charge carriers in the conducting channel is of paramount importance. Here, different approaches exist to derive this key figure for FET devices. The most commonly applied method is to measure the source-drain current $I_\m{SD}$ as a function of the gate voltage $V_\m{GS}$. Then, the field effect mobility $\mu_\m{FE}$ is determined from a tangent to the linear region of the $I_\m{SD}(V_\m{GS})$-dependence using the following formula known from FET theory:
\beq\mu_\m{FE}=\dfp{I_\m{SD}}{V_\m{GS}}\cdot\f{A}{C(V_\m{GS})}\cdot\f Lw\cdot \f 1{V_\m{SD}}\,.\label{mu-tangent-formula}\eeq
Here, $C(V_\m{GS})/A$ is the capacitance per unit area, $V_\m{SD}$ the source-drain voltage, $\dfp{I_\m{SD}}{V_\m{GS}}$ the slope of the linear region, $L$ the length and $w$ the width of the conducting channel. The intersection of the tangent with the abscissa represents the threshold voltage, $V_\m{Th}$. However, this simple FET formula (\ref{mu-tangent-formula}) assumes that the mobility is independent of the gate voltage. Moreover, the underlying parallel-plate capacitor model used to quantify the capacitance \cite{Late2012, Radisavljevic2011} assumes perfectly conducting, infinitely large plates. These assumptions may represent an oversimplification for 2D semiconductors \cite{correspondence2013Fuhrer, correspondence2013Radisavljevic}. To quantify the capacitance more precisely, Radisavljevic  and coworkers \cite{Radisavljevic2013} followed an indirect approach: the capacitance was determined from the carrier density obtained from Hall effect measurements and used in equation (\ref{mu-tangent-formula}). This helps getting more reliable capacitance values than the ones from the parallel-plate capacitor formula, but the gate voltage dependence was not investigated.\\

In this Letter, we present an easy to implement approach to determine the carrier density and carrier mobility of a $\m{MoS_2-LiNbO_3}$ FET as a function of $V_\m{GS}$. For this purpose we combine standard $I_\m{SD}-V_\m{GS}$ with $C-V_\m{GS}$. The latter probes the carrier system at the chemical potential and allows us to directly derive the carrier density as a function of $V_\m{GS}$. All experimental data is found to be in excellent agreement with an analytical model based on a 2D electron system over the entire range of $V_\m{GS}$. Most strikingly we find that for our device the values of $\mu_\m{FE}$ obtained solely from the $I_\m{SD}-V_\m{GS}$ overestimates those obtained taking into account the measure $C-V_\m{GS}$ by more than a factor of $\times 4$. \\

\begin{figure}
\includegraphics[width=0.62\linewidth]{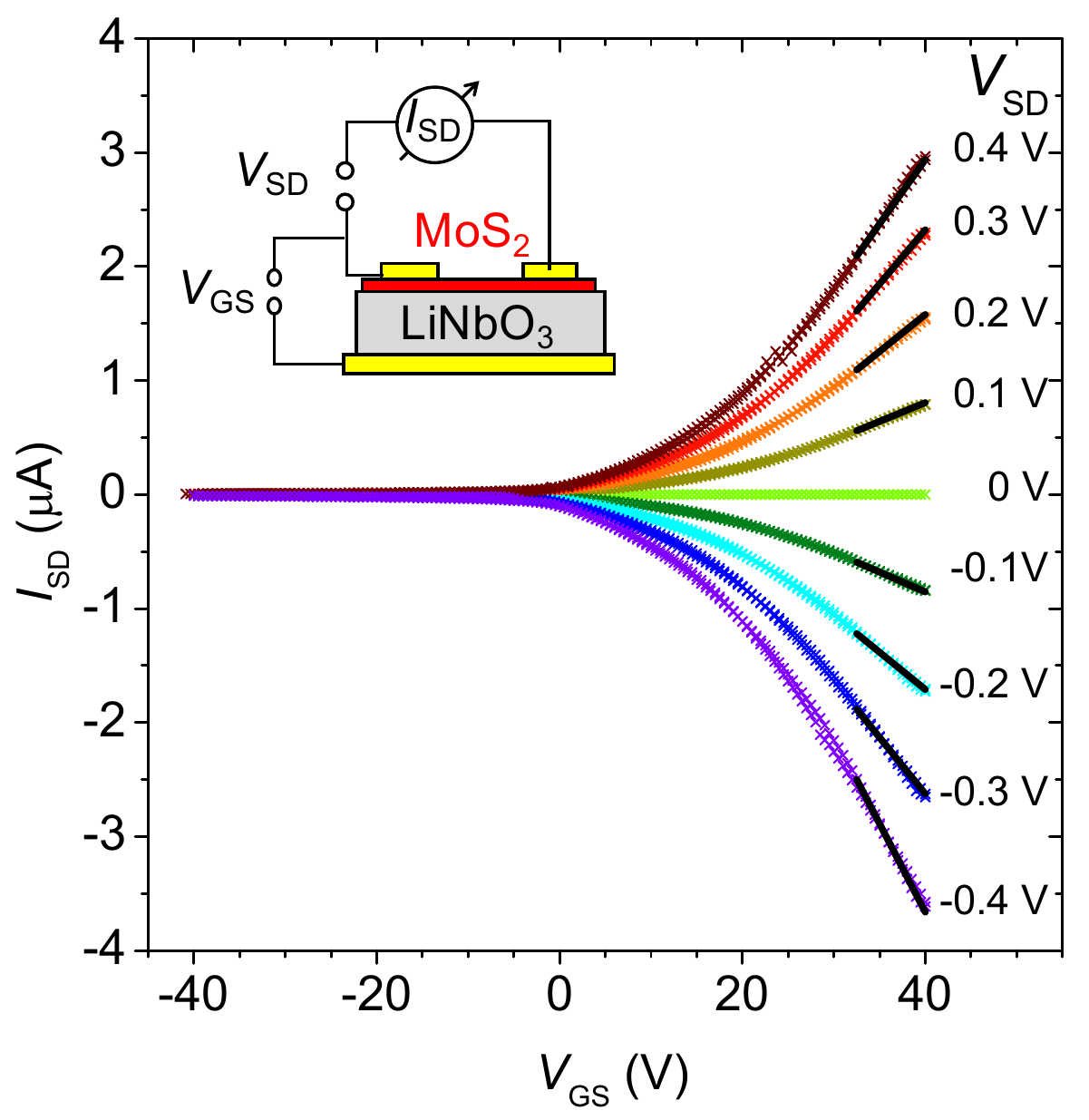}
\caption{(Color online) \label{Fig1}Measured source-drain current as function of the applied gate voltage for a set of source-drain voltages $V_\m{SD}$ tuned from $-0.4\,\m V$ (violet) to $0.4\,\m V$ (dark red) in steps of $0.1\,\m V$.  Black lines are tangent fits to the approximately linear region of the curves. Inset: Schematic of experimental setup for measuring transport characteristics with the sample, side view.}
\end{figure}

The sample studied consists of $128^\circ\mathrm{Y}$-rotated, $d=0.5\,\m{mm}$ thick substrate of black lithium niobate ($\mathrm{LiNbO}_{3-x}$), on top of which a layer of $\m{MoS_2}$ has been deposited by chemical vapour deposition \cite{Nguyen15,Preciado2015}. We use such samples for investigations of the interaction of surface acoustic waves with $\m{MoS_2}$, as described in ref. \cite{Preciado2015}.
The sample was characterized by mapping photoluminescence spectroscopy to confirm millimeter-scale growth of $\m{MoS_2}$. FET devices were fabricated using an established process\cite{Preciado2015}: the $\m{MoS_2}$ layer has been removed from the sample surface except of two $0.3\,\m{mm}\times 6.43\,\m{mm}$ stripes. This $A\approx 3.86\,\m{mm}^2$ lithographically defined area includes the regions of highest emission intensity. On top of these stripes, two finger electrodes ($10\,\m{nm}$ $\m{Ti}$ and $60\,\m{nm}$ $\m{Au}$) with a distance of $6.37\,$\textrm\textmu$\m m$ serve as source and drain contacts, contacting the two stripes along their entire $2\times 0.3\m{mm}$ width. The gate voltage $(V_\m{GS})$ was applied on the sample backside. A schematic of the sample is shown as inset in Fig. \ref{Fig1}. All measurements have been performed under ambient conditions with the device mounted inside a sealed metal chip carrier to exclude any influences of external illumination.

\begin{figure}
\includegraphics[width=0.99\linewidth]{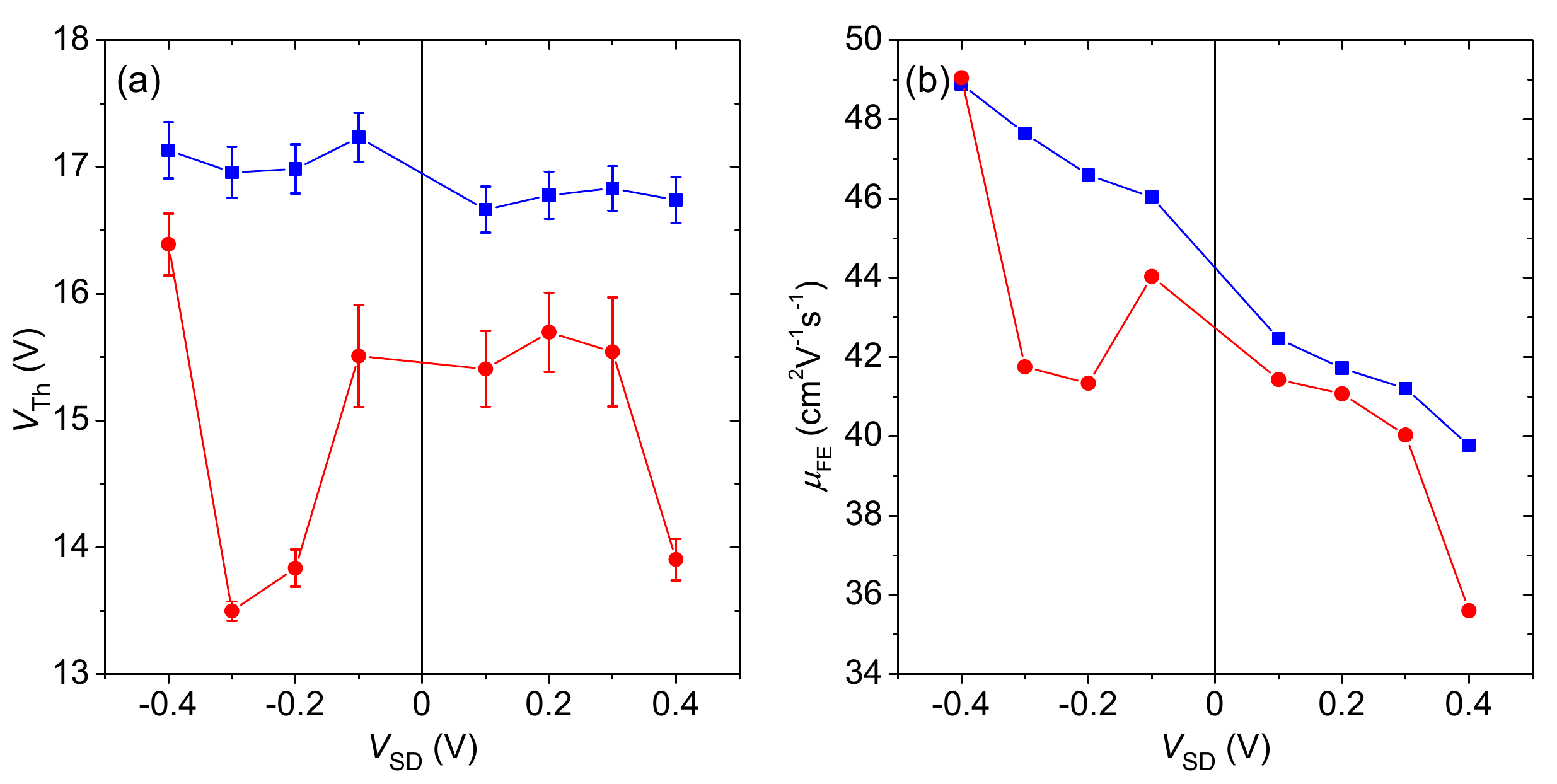}
\caption{(Color online) \label{Fig2} (a) Threshold voltage $V_\m{Th}$ and (b) the mobility $\mu_\m{FE}$ as a function of $V_\m{SD}$. Red symbols are from linear fits to the linear region of the transport characteristics shown in Fig. \ref{Fig1}. Blue symbols are extracted from the best fits of (\ref{fit-formula}) to the data in Fig. \ref{Fig1}. {The error bars represent the statistical errors of the fit.}}
\end{figure}

\begin{figure}
\includegraphics[width=0.99\columnwidth]{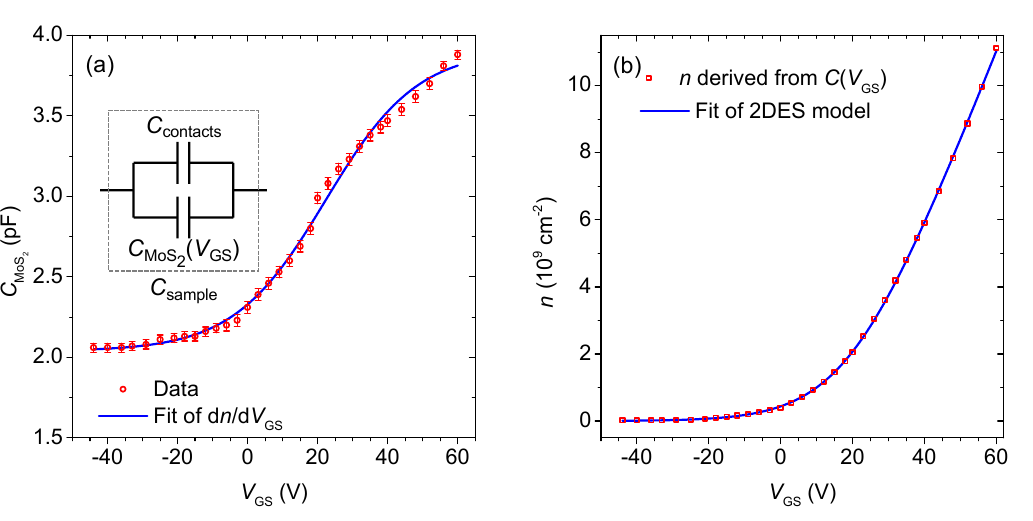}
\caption{(Color online) \label{Fig3} (a) Measured capacitance (symbols) of the sample as a function of the gate voltage, the error bars represent the statistical error and result of best fit of function (\ref{fit-C-formula}) to the data (solid lines). Inset: Equivalent circuit. (b) Electron density (symbols) obtained from the measured capacitance shifted by the constant offset and best fit by function (\ref{fit-formula}) to the data (solid line).}
\end{figure}

As a first estimate of $\mu_\m{FE}$, we measured the source-drain current $I_\m{SD}$ at a fixed source-drain voltage $V_\m{SD}$ as a function of $V_\m{GS}$, as depicted in Fig. \ref{Fig1}.
For each $V_\m{SD}$, a weak hysteresis\cite{Late2012} is resolved as the $V_\m{GS}$ was scanned $0\rightarrow+40\,\m{V}\rightarrow-40\,\m{V}\rightarrow0\,\m{V}$, which is less pronounced than that reported for a different sample in our previous work\cite{Preciado2015}.
The polarity of $V_\m{GS}$ is chosen such that $V_\m{GS}>0$ corresponds to negative charge on the $\m{MoS_2}$ layer.
These characteristics directly confirm the accumulation of negative charge on the $\m{MoS_2}$ layer for $V_\m{GS}>0$.
From this data, we extracted $V_\m{Th}$ and $\mu_\m{FE}$ using Eq. (\ref{mu-tangent-formula}) and the simple parallel-plate capacitor model.
The obtained values $V_\m{Th}$ and $\mu_\m{FE}$ are plotted as a function of $V_\m{SD}$ as red symbols in Fig. \ref{Fig2} (a) and (b), respectively.
Obviously, both the obtained values and their statistical errors exhibit significant scatter for the different values of $V_\m{SD}$.
These shortcomings arise from the simple parallel-plate capacitor model and the fact that the $V_\m{GS}$-interval, in which the best fit of Eq. (\ref{mu-tangent-formula}) is performed, is chosen by eye in the conducting region, so that all other data is neglected.
We note, that this device shows similar characteristics as that reported in our previous work\cite{Preciado2015}. In particular, $\mu_\m{FE}$ lies in the same range.
In order to improve our method, we directly quantify the capacitance between one contact and the back gate as a function of $V_\m{GS}$.
$V_\m{GS}$ is applied as a DC offset gate voltage to the capacitance bridge and modulated with a $1\,\m{kHz}$ sine wave by the built-in oscillator.
The measured capacitance was corrected for the capacitance of the wires connecting to the sample.
In Fig. \ref{Fig3} (a), we plot the obtained capacitance of the sample $C_\m{sample}$ (symbols) as a function of $V_\m{GS}$.
For large negative $V_\m{GS}$, $C_\m{sample}$ saturates at a constant value of $C_\m{sample}=2.06\,\m{pF}$.
Under these conditions, the $\m{MoS_2}$ layer is completely depleted and the measured $C_\m{sample}$ corresponds to that of the metal contacts $C_\m{contacts}$, which is independent of $V_\m{GS}$.
As $V_\m{GS}$ increases, the surrounding $\m{MoS_2}$ 2D layer is populated with electrons and the capacitance increases as observed in the data.
$C_\m{sample}(V_\m{GS})$ can be readily described as an equivalent circuit of $C_\m{contacts}$ connected in parallel with the $V_\m{GS}$-dependent capacitance of the TMD layer $C_\m{MoS_2}(V_\m{GS})$, shown as an inset of Fig. \ref{Fig3} (a).
From $C_\m{MoS_2}(V_\m{GS})=C_\m{sample}(V_\m{GS})-C_\m{contacts}$ we can directly calculate the electron surface density $n(V_\m{GS})$ on the $\m{MoS_2}$ layer by a discrete integration.
The symbols in Fig. \ref{Fig3} (b) are the result obtained from 
\[n(V_\m{GS})=-\f 1{e\cdot A}\cdot\int_{-\infty}^{V_\m{GS}}C_\m{MoS_2}(V_\m{GS}')\m dV_\m{GS}',\] with $e$ being the elementary charge.
The obtained values for $n(V_\m{GS})$ faithfully reproduce a clear turn-on behavior and linear increase as expected for a FET.\\ 

{We proceed by developing an analytical model of the $V_\m{GS}$-dependent electron density $n$. The equilibrium electron density can be calculated by integrating the Fermi distribution function $f_\m{FD}(E)=\left(\exp{\f{(E-\zeta)}{k_BT}}+1\right)^{-1}$ over the two-dimensional momenta $\v p$, 
\beq n=\frac gA\cdot\sum_{\v p}f_\m{FD}(E(\v p))=\f{2\pi gm^*\cdot k_\m BT}{h^2}\cdot\ln\lr[[]{1+\exp\lr{\f{\zeta}{k_\m BT}}}\,,\label{n-zeta}\eeq
where $g=2\cdot 2$ is the spin and valley degeneracy for $\m{MoS_2}$, $E(p)=p^2/2m^*$, $m^*$ is the electron effective mass, $A$ the area of the monolayer, and $h$ the Planck constant
; the energy is counted from the conduction band edge in MoS$_2$. In the presence of a gate voltage the chemical potential $\zeta$ in (\ref{n-zeta}) should be modified as $\zeta\to \zeta+\delta\zeta(V_\m{GS})$, where $\delta\zeta(V_\m{GS})$ is the shift of $\zeta$ in $\m{MoS}_2$ due to $V_\m{GS}$. We assume that $\delta\zeta(V_\m{GS})$ is proportional to $V_\m{GS}$, i.e., $\delta\zeta(V_\m{GS})=\alpha\cdot e\cdot V_\m{GS}$ and use the following function to fit our experimental data 
\beq
f(V_\m{GS})=a\cdot b\cdot\ln\lr[[]{1+\exp\lr{\f{V_\m{GS}-V_\m{Th}}b}}\,,\label{fit-formula}
\eeq
with $V_\m{Th}$ being a threshold voltage. 
The result of the best fit of this function to the $n$ derived from the measured capacitance is plotted as the solid line in Fig. \ref{Fig2}(b) and shows that this analytical function perfectly follows the experimental data over the entire range of $V_\m{GS}$.

The values of the parameters $a$ and $b$ extracted from the fit are $a=2.73\cdot 10^8\m{V^{-1}cm^{-2}}$ and $b=10.8\m V$. Since $k_\m BT/e\approx 26\m{mV}$ we get from $b$ the value of $\alpha\approx 2.4\cdot 10^{-3}$. Such a small value of $\alpha=\delta\zeta(V_\m{GS})/e V_\m{GS}$ suggests that the position of the electrochemical potential in the $\m{MoS}_2$--metal contacts system is essentially determined by the contacts, where the density of states is substantially larger than in $\m{MoS}_2$. Indeed, in equilibrium (without gate voltage) a small Schottky barrier, $\lesssim 0.1\,\m{eV}$, is typically built up at the metal--TMD interface \cite{Yoon2011}. Under these conditions, a certain amount of electrons flow into $\m{MoS}_2$ from the contacts, so that the density $n$ is inhomogeneous and larger in the near-contact areas than in the areas farther away. A positive $V_\m{GS}$ increases the electron density in the top metal contact. The additional electrons accumulate at the bottom of the contact in a layer with a thickness corresponding to the Thomas-Fermi screening length, what leads to a small increase of the electrochemical potential $\delta\zeta(V_\m{GS})$ in the $\m{MoS}_2$--metal contacts system. Due to the high density of states in the metal, $\delta\zeta(V_\m{GS})$ is much smaller than $e\cdot U_\m G$, which yields $\alpha\ll 1$. Due to the growth of $\zeta(V_\m{GS})$, more electrons flow into $\m{MoS}_2$ and larger areas of the 2D semiconductor become well-conducting. This basic physical picture qualitatively agrees with our results.

Comparing the equations (\ref{n-zeta}) and (\ref{fit-formula}), one can see that $a=2\pi gm^*\cdot e\alpha/h^2$. Using $\alpha$ extracted from $b$ and $m^*\approx 0.45\cdot m_\m{e,\,0}$\cite{Yoon2011}, one gets $a\approx 9\cdot 10^{11}\m{V^{-1}cm^{-2}}$. In order to explain the deviation from the fit value, a more accurate model is necessary, which takes into account the concrete contact geometry and coordinate dependence.\\
}

An analytical expression for the capacitance can be directly obtained by taking the derivative of Eq.(\ref{fit-formula}).
We obtain 
\beq C(V_\m{GS})=\lr{C_\infty-C_\m{contacts}}\cdot\f 1{\exp\lr{\f{V_\m{Th}-V_\m{GS}}{b}}}+C_\m{contacts}, \,\label{fit-C-formula}\eeq
with $C_\infty$ being the maximum capacitance for $V_\m{GS}\rightarrow\infty$.
The result of the best fit of Eq. (\ref{fit-C-formula}) to the measured capacitance is shown as a solid line in Fig. \ref{Fig3} (a) which again faithfully reproduces the experimental data points.
\\

\begin{figure}
\includegraphics[width=0.99\linewidth]{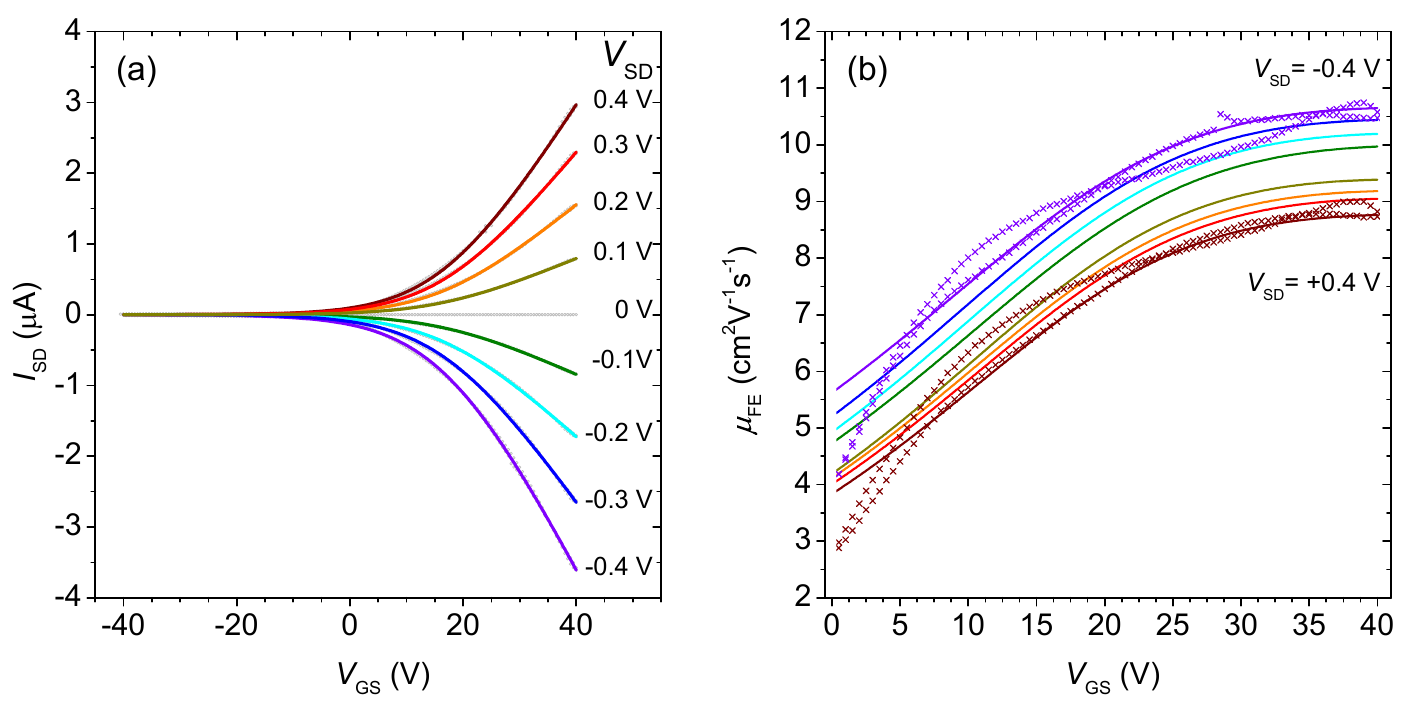}
\caption{(Color online) \label{Fig4} (a) Best fits (solid lines) of function in (\ref{fit-formula}) to measured $I_\m{SD}(V_\m{GS})$ characteristics of Fig. \ref{Fig1} (grey symbols).  
(b) Field effect mobility $\mu_\m{FE}$ as a function of $V_\m{GS}$ as obtained from the measured (symbols) and fitted (lines) $I_\m{SD}(V_\m{GS})$ and fitted $C(V_\m{GS})$ data for different values of $V_\m{SD}$. 
}
\end{figure}

In the next step, we assume that $\mu_\m{FE}$ is independent of $V_\m{GS}$. Thus, in the Drude model, $I_\m{SD}(V_\m{GS})\propto n(V_\m{GS})$ can be fitted using Eq. (\ref{fit-formula}). The results of such best fits for \emph{all} measured $I_\m{SD}(V_\m{GS})$ are plotted as solid lines in Fig. \ref{Fig4} (a). Again, the fitted function faithfully reproduces the experimental data, underlining further the 2DES nature of the conducting channel. Furthermore, these obtained fit functions allow to determine $V_\m{Th}(V_\m{SD})$ and $\mu_\m{FE}(V_\m{SD})$, using Eq. (\ref{mu-tangent-formula}) with $\partial I_\m{SD}/\partial V_\m{GS}=a$ as the slope at large $V_\m{GS}$, with higher precision. The extracted values for $V_\m{Th}(V_\m{SD})$ and $\mu_\m{FE}(V_\m{SD})$ are plotted as blue symbols in Fig. \ref{Fig2} (a) and (b), respectively. Clearly, the scatter of the values derived from the fit results is dramatically reduced. We obtain $\langle V_\m{Th}\rangle =(17.0\pm0.2)\m{V}$, which is almost constant over the entire range of $V_\m{SD}$. 
In contrast, $\mu_\m{FE}(V_\m{SD})$ exhibits a clear trend to significantly decrease for increasing $V_\m{SD}$. The negative slope of the $V_\m{SD}$-dependence of the mobility has its reason in  {predominantly in hysteresis and drifts of the electrical characteristics.
Both effects are commonly observed in such devices\cite{Late2012,Illarionov2016}.
}

Finally, we turn to carrier mobility and its dependence on the gate voltage, $\mu_\m{FE}(V_\m{GS})$. 
In the Drude model the conductivity is given by $\sigma=e\cdot n\cdot\mu_\m{FE}$.
Thus, the mobility given by $\mu_\m{FE}(V_\m{GS})=\sigma(V_\m{GS})/(e\cdot n(V_\m{GS}))$ can be calculated only from measured data: $\sigma(V_\m{GS})$ can be derived from the $I_\m{SD}(V_\m{GS})$ characteristics [Figs. \ref{Fig1} and \ref{Fig4} (a)] and $n(V_\m{GS})$ from the $C_\m{MoS_2}(V_\m{GS})$ data [\textit{cf.} Fig. \ref{Fig4} (b)].
We note that this analysis can be performed for our data only for $V_\m{GS}\gtrsim 0\,\m V$. For negative $V_\m{GS}$ both $\sigma$ and $n$ vanish and any obtained value of $\mu_\m{FE}\sim \sigma/n$ exhibits a large error.
In Fig. \ref{Fig2} (b) we plot $\mu_\m{FE}(V_\m{GS})$ obtained directly from the measured (symbols) and the fitted (lines) $I_\m{SD}(V_\m{GS})$ characteristics for different values of $V_\m{SD}$ for $V_\m{GS}>0\,\m V$.
Remarkably, the \emph{absolute} value of $\mu_\m{FE}$ shown in Fig. \ref{Fig4} (b), which we obtained by including \emph{measured, realistic} capacitance (and thereby $n$) data is significantly lower than that obtained from the basic parallel-plate capacitor model [\textit{cf.} Fig. \ref{Fig2} (b)].
It ranges between $\mu_\m{FE}(V_\m{GS}) \sim 4 - 10\,\m{cm^2V^{-1}s^{-1}}$ and shows a pronounced increase with increasing $V_\m{GS}$ (in addition to its global reduction as $V_\m{SD}$ reduces [cf. Fig. \ref{Fig2} (b)]).
For $0<V_\m{GS}<V_\m{Th}$, $\mu_\m{FE}$ rapidly increases as the injected electrons screen scattering centers in the channel.
Such behavior is well established and has been observed for 2DES in established III-V semiconductor heterostructures\cite{Shayegan1988,Pfeiffer1989}.
For $V_\m{GS}>V_\m{Th}$ this trend weakens and $\mu_\m{FE}$ saturates.
This saturation behavior can be readily understood considering that as the chemical potential $\zeta$ is fully shifted into the conduction band.
For large positive $V_\m{GS}$ the chemical potential lies well above the random potential modulation induced by scattering centers.
Thus, any further increase of $\zeta$ (i.e. $n$) does not lead to improved screening and, thus increased $\mu_\m{FE}$.\\

In summary, we demonstrated that combined electrical transport ($I_\m{SD}-V_\m{GS}$) and capacitance ($C-V_\m{GS}$) spectroscopy allows to determine the field-effect mobility $\mu_\m{FE}$ and threshold voltage $V_\m{Th}$ of a TMD based FET with significantly higher precision than the commonly applied basic parallel-plate capacitor model.
We performed a three step analysis on model data of a $\m{MoS_2}$-$\m{LiNbO_3}$ FET device starting with the basic parallel-plate capacitor model.
For our device the $V_\m{GS}$-dependent $I_\m{SD}$, $n$ and $C$ are in excellent agreement with an analytical model of an ideal 2DES over the \emph{entire range of} $V_\m{GS}$. 
This is in strong contrast to the basic parallel-plate capacitor model in which only data in a small, subjectively chosen interval of the $I_\m{SD}-V_\m{GS}$-characteristics is considered.
The statistical errors of $\mu_\m{FE}$ and $V_\m{Th}$ can be significantly reduced by fitting and evaluating the $I_\m{SD}-V_\m{GS}$-characteristics using our 2DES model as now the full data range is included.
Finally, by including the full $C-V_\m{GS}$-characteristics and the derived carrier density $n$ we are able to obtain the $V_\m{GS}$-dependent $\mu_\m{FE}$.
For our device we nicely observe a pronounced increase of $\mu_\m{FE}$ at $V_\m{Th}$ due to onset of efficient electrostatic screening.
Most strikingly, the absolute value of $\mu_\m{FE}$ obtained this way is significantly lower than that obtained from the basic parallel-plate capacitor model.
Our full method is especially important for back-gated 2D material-based FETs.
For such devices, the distance between the 2D carrier system and the gate electrode is large compared to the lateral dimensions of typical flakes.
It enables (i) direct confirmation of 2DES-like character of the FET operation, and derive (ii) more accurate, realistic and (iii) $V_\m{GS}$-dependent values $\mu_\m{FE}$. Moreover, it does not require a magnetic field as the mobility determination via Hall effect, thus making it suitable for setups without magnets and for samples with low mobilities, for which the Hall angle is small.\\

This work was supported by the Deutsche Forschungsgemeinschaft (DFG) via the Emmy Noether Program (KR3790/2), by the Cluster of Excellence "Nanosystems Initiative Munich" (NIM), by the Bavaria-California Technology Center (BaCaTeC). L.B. thanks the National Science Foundation for support under NSF DMR 1609918 and C-SPIN, a STARnet center funded by MARCO and DARPA. E.P. and A.E.N. and G.S. gratefully acknowledge fellowship support through NSF DGE-1326120 and NSF DMR 1359136, respectively.


\begin{thebibliography}{22}%
\makeatletter
\providecommand \@ifxundefined [1]{%
 \@ifx{#1\undefined}
}%
\providecommand \@ifnum [1]{%
 \ifnum #1\expandafter \@firstoftwo
 \else \expandafter \@secondoftwo
 \fi
}%
\providecommand \@ifx [1]{%
 \ifx #1\expandafter \@firstoftwo
 \else \expandafter \@secondoftwo
 \fi
}%
\providecommand \natexlab [1]{#1}%
\providecommand \enquote  [1]{``#1''}%
\providecommand \bibnamefont  [1]{#1}%
\providecommand \bibfnamefont [1]{#1}%
\providecommand \citenamefont [1]{#1}%
\providecommand \href@noop [0]{\@secondoftwo}%
\providecommand \href [0]{\begingroup \@sanitize@url \@href}%
\providecommand \@href[1]{\@@startlink{#1}\@@href}%
\providecommand \@@href[1]{\endgroup#1\@@endlink}%
\providecommand \@sanitize@url [0]{\catcode `\\12\catcode `\$12\catcode
  `\&12\catcode `\#12\catcode `\^12\catcode `\_12\catcode `\%12\relax}%
\providecommand \@@startlink[1]{}%
\providecommand \@@endlink[0]{}%
\providecommand \url  [0]{\begingroup\@sanitize@url \@url }%
\providecommand \@url [1]{\endgroup\@href {#1}{\urlprefix }}%
\providecommand \urlprefix  [0]{URL }%
\providecommand \Eprint [0]{\href }%
\providecommand \doibase [0]{http://dx.doi.org/}%
\providecommand \selectlanguage [0]{\@gobble}%
\providecommand \bibinfo  [0]{\@secondoftwo}%
\providecommand \bibfield  [0]{\@secondoftwo}%
\providecommand \translation [1]{[#1]}%
\providecommand \BibitemOpen [0]{}%
\providecommand \bibitemStop [0]{}%
\providecommand \bibitemNoStop [0]{.\EOS\space}%
\providecommand \EOS [0]{\spacefactor3000\relax}%
\providecommand \BibitemShut  [1]{\csname bibitem#1\endcsname}%
\let\auto@bib@innerbib\@empty
\bibitem [{\citenamefont {Novoselov}\ \emph {et~al.}(2004)\citenamefont
  {Novoselov}, \citenamefont {Geim}, \citenamefont {Morozov}, \citenamefont
  {Jiang}, \citenamefont {Zhang}, \citenamefont {Dubonos}, \citenamefont
  {Grigorieva},\ and\ \citenamefont {Firsov}}]{Novoselov04}%
  \BibitemOpen
  \bibfield  {author} {\bibinfo {author} {\bibfnamefont {K.~S.}\ \bibnamefont
  {Novoselov}}, \bibinfo {author} {\bibfnamefont {A.~K.}\ \bibnamefont {Geim}},
  \bibinfo {author} {\bibfnamefont {S.~V.}\ \bibnamefont {Morozov}}, \bibinfo
  {author} {\bibfnamefont {D.}~\bibnamefont {Jiang}}, \bibinfo {author}
  {\bibfnamefont {Y.}~\bibnamefont {Zhang}}, \bibinfo {author} {\bibfnamefont
  {S.~V.}\ \bibnamefont {Dubonos}}, \bibinfo {author} {\bibfnamefont {I.~V.}\
  \bibnamefont {Grigorieva}}, \ and\ \bibinfo {author} {\bibfnamefont {A.~A.}\
  \bibnamefont {Firsov}},\ }\bibfield  {title} {\enquote {\bibinfo {title}
  {Electric field effect in atomically thin carbon films},}\ }\href@noop {}
  {\bibfield  {journal} {\bibinfo  {journal} {Science}\ }\textbf {\bibinfo
  {volume} {306}},\ \bibinfo {pages} {666--669} (\bibinfo {year}
  {2004})}\BibitemShut {NoStop}%
\bibitem [{\citenamefont {Novoselov}\ \emph
  {et~al.}(2005{\natexlab{a}})\citenamefont {Novoselov}, \citenamefont {Geim},
  \citenamefont {Morozov}, \citenamefont {Jiang}, \citenamefont {Katsnelson},
  \citenamefont {Grigorieva}, \citenamefont {Dubonos},\ and\ \citenamefont
  {Firsov}}]{Novoselov05}%
  \BibitemOpen
  \bibfield  {author} {\bibinfo {author} {\bibfnamefont {K.~S.}\ \bibnamefont
  {Novoselov}}, \bibinfo {author} {\bibfnamefont {A.~K.}\ \bibnamefont {Geim}},
  \bibinfo {author} {\bibfnamefont {S.~V.}\ \bibnamefont {Morozov}}, \bibinfo
  {author} {\bibfnamefont {D.}~\bibnamefont {Jiang}}, \bibinfo {author}
  {\bibfnamefont {M.~I.}\ \bibnamefont {Katsnelson}}, \bibinfo {author}
  {\bibfnamefont {I.~V.}\ \bibnamefont {Grigorieva}}, \bibinfo {author}
  {\bibfnamefont {S.~V.}\ \bibnamefont {Dubonos}}, \ and\ \bibinfo {author}
  {\bibfnamefont {A.~A.}\ \bibnamefont {Firsov}},\ }\bibfield  {title}
  {\enquote {\bibinfo {title} {Two-dimensional gas of massless {D}irac fermions
  in graphene},}\ }\href@noop {} {\bibfield  {journal} {\bibinfo  {journal}
  {Nature}\ }\textbf {\bibinfo {volume} {438}},\ \bibinfo {pages} {197--200}
  (\bibinfo {year} {2005}{\natexlab{a}})}\BibitemShut {NoStop}%
\bibitem [{\citenamefont {Zhang}\ \emph {et~al.}(2005)\citenamefont {Zhang},
  \citenamefont {Tan}, \citenamefont {Stormer},\ and\ \citenamefont
  {Kim}}]{Zhang05}%
  \BibitemOpen
  \bibfield  {author} {\bibinfo {author} {\bibfnamefont {Y.}~\bibnamefont
  {Zhang}}, \bibinfo {author} {\bibfnamefont {Y.-W.}\ \bibnamefont {Tan}},
  \bibinfo {author} {\bibfnamefont {H.~L.}\ \bibnamefont {Stormer}}, \ and\
  \bibinfo {author} {\bibfnamefont {P.}~\bibnamefont {Kim}},\ }\bibfield
  {title} {\enquote {\bibinfo {title} {Experimental observation of the quantum
  {H}all effect and {B}erry's phase in graphene},}\ }\href@noop {} {\bibfield
  {journal} {\bibinfo  {journal} {Nature}\ }\textbf {\bibinfo {volume} {438}},\
  \bibinfo {pages} {201--204} (\bibinfo {year} {2005})}\BibitemShut {NoStop}%
\bibitem [{\citenamefont {Novoselov}\ \emph
  {et~al.}(2005{\natexlab{b}})\citenamefont {Novoselov}, \citenamefont {Jiang},
  \citenamefont {Schedin}, \citenamefont {Booth}, \citenamefont {Khotkevich},
  \citenamefont {Morozov},\ and\ \citenamefont {Geim}}]{Novoselov05-2D}%
  \BibitemOpen
  \bibfield  {author} {\bibinfo {author} {\bibfnamefont {K.~S.}\ \bibnamefont
  {Novoselov}}, \bibinfo {author} {\bibfnamefont {D.}~\bibnamefont {Jiang}},
  \bibinfo {author} {\bibfnamefont {F.}~\bibnamefont {Schedin}}, \bibinfo
  {author} {\bibfnamefont {T.~J.}\ \bibnamefont {Booth}}, \bibinfo {author}
  {\bibfnamefont {V.~V.}\ \bibnamefont {Khotkevich}}, \bibinfo {author}
  {\bibfnamefont {S.~V.}\ \bibnamefont {Morozov}}, \ and\ \bibinfo {author}
  {\bibfnamefont {A.~K.}\ \bibnamefont {Geim}},\ }\bibfield  {title} {\enquote
  {\bibinfo {title} {Two-dimensional atomic crystals},}\ }\href@noop {}
  {\bibfield  {journal} {\bibinfo  {journal} {Proceedings of the National
  Academy of Sciences of the United States of America}\ }\textbf {\bibinfo
  {volume} {102}},\ \bibinfo {pages} {10451--10453} (\bibinfo {year}
  {2005}{\natexlab{b}})}\BibitemShut {NoStop}%
\bibitem [{\citenamefont {Ferrari}\ \emph {et~al.}(2015)\citenamefont
  {Ferrari}, \citenamefont {Bonaccorso}, \citenamefont {Fal{'}ko},
  \citenamefont {Novoselov}, \citenamefont {Roche}, \citenamefont {Boggild},
  \citenamefont {Borini}, \citenamefont {Koppens}, \citenamefont {Palermo},
  \citenamefont {Pugno}, \citenamefont {Garrido}, \citenamefont {Sordan},
  \citenamefont {Bianco}, \citenamefont {Ballerini}, \citenamefont {Prato},
  \citenamefont {Lidorikis}, \citenamefont {Kivioja}, \citenamefont
  {Marinelli}, \citenamefont {Ryhanen}, \citenamefont {Morpurgo}, \citenamefont
  {Coleman}, \citenamefont {Nicolosi}, \citenamefont {Colombo}, \citenamefont
  {Fert}, \citenamefont {Garcia-Hernandez}, \citenamefont {Bachtold},
  \citenamefont {Schneider}, \citenamefont {Guinea}, \citenamefont {Dekker},
  \citenamefont {Barbone}, \citenamefont {Sun}, \citenamefont {Galiotis},
  \citenamefont {Grigorenko}, \citenamefont {Konstantatos}, \citenamefont
  {Kis}, \citenamefont {Katsnelson}, \citenamefont {Vandersypen}, \citenamefont
  {Loiseau}, \citenamefont {Morandi}, \citenamefont {Neumaier}, \citenamefont
  {Treossi}, \citenamefont {Pellegrini}, \citenamefont {Polini}, \citenamefont
  {Tredicucci}, \citenamefont {Williams}, \citenamefont {Hee~Hong},
  \citenamefont {Ahn}, \citenamefont {Min~Kim}, \citenamefont {Zirath},
  \citenamefont {van Wees}, \citenamefont {van~der Zant}, \citenamefont
  {Occhipinti}, \citenamefont {Di~Matteo}, \citenamefont {Kinloch},
  \citenamefont {Seyller}, \citenamefont {Quesnel}, \citenamefont {Feng},
  \citenamefont {Teo}, \citenamefont {Rupesinghe}, \citenamefont {Hakonen},
  \citenamefont {Neil}, \citenamefont {Tannock}, \citenamefont {Lofwander},\
  and\ \citenamefont {Kinaret}}]{roadmap-graphene}%
  \BibitemOpen
  \bibfield  {author} {\bibinfo {author} {\bibfnamefont {A.~C.}\ \bibnamefont
  {Ferrari}}, \bibinfo {author} {\bibfnamefont {F.}~\bibnamefont {Bonaccorso}},
  \bibinfo {author} {\bibfnamefont {V.}~\bibnamefont {Fal{'}ko}}, \bibinfo
  {author} {\bibfnamefont {K.~S.}\ \bibnamefont {Novoselov}}, \bibinfo {author}
  {\bibfnamefont {S.}~\bibnamefont {Roche}}, \bibinfo {author} {\bibfnamefont
  {P.}~\bibnamefont {Boggild}}, \bibinfo {author} {\bibfnamefont
  {S.}~\bibnamefont {Borini}}, \bibinfo {author} {\bibfnamefont {F.~H.~L.}\
  \bibnamefont {Koppens}}, \bibinfo {author} {\bibfnamefont {V.}~\bibnamefont
  {Palermo}}, \bibinfo {author} {\bibfnamefont {N.}~\bibnamefont {Pugno}},
  \bibinfo {author} {\bibfnamefont {J.~A.}\ \bibnamefont {Garrido}}, \bibinfo
  {author} {\bibfnamefont {R.}~\bibnamefont {Sordan}}, \bibinfo {author}
  {\bibfnamefont {A.}~\bibnamefont {Bianco}}, \bibinfo {author} {\bibfnamefont
  {L.}~\bibnamefont {Ballerini}}, \bibinfo {author} {\bibfnamefont
  {M.}~\bibnamefont {Prato}}, \bibinfo {author} {\bibfnamefont
  {E.}~\bibnamefont {Lidorikis}}, \bibinfo {author} {\bibfnamefont
  {J.}~\bibnamefont {Kivioja}}, \bibinfo {author} {\bibfnamefont
  {C.}~\bibnamefont {Marinelli}}, \bibinfo {author} {\bibfnamefont
  {T.}~\bibnamefont {Ryhanen}}, \bibinfo {author} {\bibfnamefont
  {A.}~\bibnamefont {Morpurgo}}, \bibinfo {author} {\bibfnamefont {J.~N.}\
  \bibnamefont {Coleman}}, \bibinfo {author} {\bibfnamefont {V.}~\bibnamefont
  {Nicolosi}}, \bibinfo {author} {\bibfnamefont {L.}~\bibnamefont {Colombo}},
  \bibinfo {author} {\bibfnamefont {A.}~\bibnamefont {Fert}}, \bibinfo {author}
  {\bibfnamefont {M.}~\bibnamefont {Garcia-Hernandez}}, \bibinfo {author}
  {\bibfnamefont {A.}~\bibnamefont {Bachtold}}, \bibinfo {author}
  {\bibfnamefont {G.~F.}\ \bibnamefont {Schneider}}, \bibinfo {author}
  {\bibfnamefont {F.}~\bibnamefont {Guinea}}, \bibinfo {author} {\bibfnamefont
  {C.}~\bibnamefont {Dekker}}, \bibinfo {author} {\bibfnamefont
  {M.}~\bibnamefont {Barbone}}, \bibinfo {author} {\bibfnamefont
  {Z.}~\bibnamefont {Sun}}, \bibinfo {author} {\bibfnamefont {C.}~\bibnamefont
  {Galiotis}}, \bibinfo {author} {\bibfnamefont {A.~N.}\ \bibnamefont
  {Grigorenko}}, \bibinfo {author} {\bibfnamefont {G.}~\bibnamefont
  {Konstantatos}}, \bibinfo {author} {\bibfnamefont {A.}~\bibnamefont {Kis}},
  \bibinfo {author} {\bibfnamefont {M.}~\bibnamefont {Katsnelson}}, \bibinfo
  {author} {\bibfnamefont {L.}~\bibnamefont {Vandersypen}}, \bibinfo {author}
  {\bibfnamefont {A.}~\bibnamefont {Loiseau}}, \bibinfo {author} {\bibfnamefont
  {V.}~\bibnamefont {Morandi}}, \bibinfo {author} {\bibfnamefont
  {D.}~\bibnamefont {Neumaier}}, \bibinfo {author} {\bibfnamefont
  {E.}~\bibnamefont {Treossi}}, \bibinfo {author} {\bibfnamefont
  {V.}~\bibnamefont {Pellegrini}}, \bibinfo {author} {\bibfnamefont
  {M.}~\bibnamefont {Polini}}, \bibinfo {author} {\bibfnamefont
  {A.}~\bibnamefont {Tredicucci}}, \bibinfo {author} {\bibfnamefont {G.~M.}\
  \bibnamefont {Williams}}, \bibinfo {author} {\bibfnamefont {B.}~\bibnamefont
  {Hee~Hong}}, \bibinfo {author} {\bibfnamefont {J.-H.}\ \bibnamefont {Ahn}},
  \bibinfo {author} {\bibfnamefont {J.}~\bibnamefont {Min~Kim}}, \bibinfo
  {author} {\bibfnamefont {H.}~\bibnamefont {Zirath}}, \bibinfo {author}
  {\bibfnamefont {B.~J.}\ \bibnamefont {van Wees}}, \bibinfo {author}
  {\bibfnamefont {H.}~\bibnamefont {van~der Zant}}, \bibinfo {author}
  {\bibfnamefont {L.}~\bibnamefont {Occhipinti}}, \bibinfo {author}
  {\bibfnamefont {A.}~\bibnamefont {Di~Matteo}}, \bibinfo {author}
  {\bibfnamefont {I.~A.}\ \bibnamefont {Kinloch}}, \bibinfo {author}
  {\bibfnamefont {T.}~\bibnamefont {Seyller}}, \bibinfo {author} {\bibfnamefont
  {E.}~\bibnamefont {Quesnel}}, \bibinfo {author} {\bibfnamefont
  {X.}~\bibnamefont {Feng}}, \bibinfo {author} {\bibfnamefont {K.}~\bibnamefont
  {Teo}}, \bibinfo {author} {\bibfnamefont {N.}~\bibnamefont {Rupesinghe}},
  \bibinfo {author} {\bibfnamefont {P.}~\bibnamefont {Hakonen}}, \bibinfo
  {author} {\bibfnamefont {S.~R.~T.}\ \bibnamefont {Neil}}, \bibinfo {author}
  {\bibfnamefont {Q.}~\bibnamefont {Tannock}}, \bibinfo {author} {\bibfnamefont
  {T.}~\bibnamefont {Lofwander}}, \ and\ \bibinfo {author} {\bibfnamefont
  {J.}~\bibnamefont {Kinaret}},\ }\bibfield  {title} {\enquote {\bibinfo
  {title} {Science and technology roadmap for graphene{,} related
  two-dimensional crystals{,} and hybrid systems},}\ }\href@noop {} {\bibfield
  {journal} {\bibinfo  {journal} {Nanoscale}\ }\textbf {\bibinfo {volume}
  {7}},\ \bibinfo {pages} {4598--4810} (\bibinfo {year} {2015})}\BibitemShut
  {NoStop}%
\bibitem [{\citenamefont {Wang}\ \emph {et~al.}(2012)\citenamefont {Wang},
  \citenamefont {Kalantar-Zadeh}, \citenamefont {Kis}, \citenamefont
  {Coleman},\ and\ \citenamefont {Strano}}]{overview-article}%
  \BibitemOpen
  \bibfield  {author} {\bibinfo {author} {\bibfnamefont {Q.~H.}\ \bibnamefont
  {Wang}}, \bibinfo {author} {\bibfnamefont {K.}~\bibnamefont
  {Kalantar-Zadeh}}, \bibinfo {author} {\bibfnamefont {A.}~\bibnamefont {Kis}},
  \bibinfo {author} {\bibfnamefont {J.~N.}\ \bibnamefont {Coleman}}, \ and\
  \bibinfo {author} {\bibfnamefont {M.~S.}\ \bibnamefont {Strano}},\ }\bibfield
   {title} {\enquote {\bibinfo {title} {Electronics and optoelectronics of
  two-dimensional transition metal dichalcogenides},}\ }\href@noop {}
  {\bibfield  {journal} {\bibinfo  {journal} {Nature Nanotechnology}\ }\textbf
  {\bibinfo {volume} {7}},\ \bibinfo {pages} {699--712} (\bibinfo {year}
  {2012})}\BibitemShut {NoStop}%
\bibitem [{\citenamefont {Mak}\ \emph {et~al.}(2010)\citenamefont {Mak},
  \citenamefont {Lee}, \citenamefont {Hone}, \citenamefont {Shan},\ and\
  \citenamefont {Heinz}}]{mos2-bandgap2}%
  \BibitemOpen
  \bibfield  {author} {\bibinfo {author} {\bibfnamefont {K.~F.}\ \bibnamefont
  {Mak}}, \bibinfo {author} {\bibfnamefont {C.}~\bibnamefont {Lee}}, \bibinfo
  {author} {\bibfnamefont {J.}~\bibnamefont {Hone}}, \bibinfo {author}
  {\bibfnamefont {J.}~\bibnamefont {Shan}}, \ and\ \bibinfo {author}
  {\bibfnamefont {T.~F.}\ \bibnamefont {Heinz}},\ }\bibfield  {title} {\enquote
  {\bibinfo {title} {Atomically thin $\m{MoS}_{2}$: A new direct-gap
  semiconductor},}\ }\href@noop {} {\bibfield  {journal} {\bibinfo  {journal}
  {Physical Review Letters}\ }\textbf {\bibinfo {volume} {105}},\ \bibinfo
  {pages} {136805} (\bibinfo {year} {2010})}\BibitemShut {NoStop}%
\bibitem [{\citenamefont {Splendiani}\ \emph {et~al.}(2010)\citenamefont
  {Splendiani}, \citenamefont {Sun}, \citenamefont {Zhang}, \citenamefont {Li},
  \citenamefont {Kim}, \citenamefont {Chim}, \citenamefont {Galli},\ and\
  \citenamefont {Wang}}]{Splendiani2010}%
  \BibitemOpen
  \bibfield  {author} {\bibinfo {author} {\bibfnamefont {A.}~\bibnamefont
  {Splendiani}}, \bibinfo {author} {\bibfnamefont {L.}~\bibnamefont {Sun}},
  \bibinfo {author} {\bibfnamefont {Y.}~\bibnamefont {Zhang}}, \bibinfo
  {author} {\bibfnamefont {T.}~\bibnamefont {Li}}, \bibinfo {author}
  {\bibfnamefont {J.}~\bibnamefont {Kim}}, \bibinfo {author} {\bibfnamefont
  {C.-Y.}\ \bibnamefont {Chim}}, \bibinfo {author} {\bibfnamefont
  {G.}~\bibnamefont {Galli}}, \ and\ \bibinfo {author} {\bibfnamefont
  {F.}~\bibnamefont {Wang}},\ }\bibfield  {title} {\enquote {\bibinfo {title}
  {{Emerging photoluminescence in monolayer MoS$_2$.}}}\ }\href {\doibase
  10.1021/nl903868w} {\bibfield  {journal} {\bibinfo  {journal} {Nano Letters}\
  }\textbf {\bibinfo {volume} {10}},\ \bibinfo {pages} {1271--5} (\bibinfo
  {year} {2010})}\BibitemShut {NoStop}%
\bibitem [{\citenamefont {Lopez-Sanchez}\ \emph {et~al.}(2013)\citenamefont
  {Lopez-Sanchez}, \citenamefont {Lembke}, \citenamefont {Kayci}, \citenamefont
  {Radenovic},\ and\ \citenamefont {Kis}}]{Lopez-Sanchez2013}%
  \BibitemOpen
  \bibfield  {author} {\bibinfo {author} {\bibfnamefont {O.}~\bibnamefont
  {Lopez-Sanchez}}, \bibinfo {author} {\bibfnamefont {D.}~\bibnamefont
  {Lembke}}, \bibinfo {author} {\bibfnamefont {M.}~\bibnamefont {Kayci}},
  \bibinfo {author} {\bibfnamefont {A.}~\bibnamefont {Radenovic}}, \ and\
  \bibinfo {author} {\bibfnamefont {A.}~\bibnamefont {Kis}},\ }\bibfield
  {title} {\enquote {\bibinfo {title} {{Ultrasensitive photodetectors based on
  monolayer MoS$_2$}},}\ }\href {\doibase 10.1038/nnano.2013.100} {\bibfield
  {journal} {\bibinfo  {journal} {Nature Nanotechnology}\ }\textbf {\bibinfo
  {volume} {8}},\ \bibinfo {pages} {497--501} (\bibinfo {year}
  {2013})}\BibitemShut {NoStop}%
\bibitem [{\citenamefont {Sun}, \citenamefont {Martinez},\ and\ \citenamefont
  {Wang}(2016)}]{Sun2016}%
  \BibitemOpen
  \bibfield  {author} {\bibinfo {author} {\bibfnamefont {Z.}~\bibnamefont
  {Sun}}, \bibinfo {author} {\bibfnamefont {A.}~\bibnamefont {Martinez}}, \
  and\ \bibinfo {author} {\bibfnamefont {F.}~\bibnamefont {Wang}},\ }\bibfield
  {title} {\enquote {\bibinfo {title} {{Optical modulators with 2D layered
  materials}},}\ }\href {\doibase 10.1038/nphoton.2016.15} {\bibfield
  {journal} {\bibinfo  {journal} {Nature Photonics}\ }\textbf {\bibinfo
  {volume} {10}},\ \bibinfo {pages} {227--238} (\bibinfo {year}
  {2016})}\BibitemShut {NoStop}%
\bibitem [{\citenamefont {Radisavljevic}\ \emph {et~al.}(2011)\citenamefont
  {Radisavljevic}, \citenamefont {Radenovic}, \citenamefont {Brivio},
  \citenamefont {Giacometti},\ and\ \citenamefont {Kis}}]{Radisavljevic2011}%
  \BibitemOpen
  \bibfield  {author} {\bibinfo {author} {\bibfnamefont {B.}~\bibnamefont
  {Radisavljevic}}, \bibinfo {author} {\bibfnamefont {A.}~\bibnamefont
  {Radenovic}}, \bibinfo {author} {\bibfnamefont {J.}~\bibnamefont {Brivio}},
  \bibinfo {author} {\bibfnamefont {V.}~\bibnamefont {Giacometti}}, \ and\
  \bibinfo {author} {\bibfnamefont {A.}~\bibnamefont {Kis}},\ }\bibfield
  {title} {\enquote {\bibinfo {title} {Single-layer $\m{MoS}_2$ transistors},}\
  }\href@noop {} {\bibfield  {journal} {\bibinfo  {journal} {Nature
  Nanotechnology}\ }\textbf {\bibinfo {volume} {6}},\ \bibinfo {pages}
  {147--150} (\bibinfo {year} {2011})}\BibitemShut {NoStop}%
\bibitem [{\citenamefont {Radisavljevic}, \citenamefont {Whitwick},\ and\
  \citenamefont {Kis}(2011)}]{Radisavljevic2011Integrated}%
  \BibitemOpen
  \bibfield  {author} {\bibinfo {author} {\bibfnamefont {B.}~\bibnamefont
  {Radisavljevic}}, \bibinfo {author} {\bibfnamefont {M.~B.}\ \bibnamefont
  {Whitwick}}, \ and\ \bibinfo {author} {\bibfnamefont {A.}~\bibnamefont
  {Kis}},\ }\bibfield  {title} {\enquote {\bibinfo {title} {Integrated circuits
  and logic operations based on single-layer $\mathrm{MoS_2}$},}\ }\href@noop
  {} {\bibfield  {journal} {\bibinfo  {journal} {ACS nano}\ }\textbf {\bibinfo
  {volume} {5}},\ \bibinfo {pages} {9934--9938} (\bibinfo {year}
  {2011})}\BibitemShut {NoStop}%
\bibitem [{\citenamefont {Fiori}\ \emph {et~al.}(2014)\citenamefont {Fiori},
  \citenamefont {Bonaccorso}, \citenamefont {Iannaccone}, \citenamefont
  {Palacios}, \citenamefont {Neumaier}, \citenamefont {Seabaugh}, \citenamefont
  {Banerjee},\ and\ \citenamefont {Colombo}}]{Fiori2014}%
  \BibitemOpen
  \bibfield  {author} {\bibinfo {author} {\bibfnamefont {G.}~\bibnamefont
  {Fiori}}, \bibinfo {author} {\bibfnamefont {F.}~\bibnamefont {Bonaccorso}},
  \bibinfo {author} {\bibfnamefont {G.}~\bibnamefont {Iannaccone}}, \bibinfo
  {author} {\bibfnamefont {T.}~\bibnamefont {Palacios}}, \bibinfo {author}
  {\bibfnamefont {D.}~\bibnamefont {Neumaier}}, \bibinfo {author}
  {\bibfnamefont {A.}~\bibnamefont {Seabaugh}}, \bibinfo {author}
  {\bibfnamefont {S.~K.}\ \bibnamefont {Banerjee}}, \ and\ \bibinfo {author}
  {\bibfnamefont {L.}~\bibnamefont {Colombo}},\ }\bibfield  {title} {\enquote
  {\bibinfo {title} {{Electronics based on two-dimensional materials}},}\
  }\href {\doibase 10.1038/nnano.2014.207} {\bibfield  {journal} {\bibinfo
  {journal} {Nature Nanotechnology}\ }\textbf {\bibinfo {volume} {9}},\
  \bibinfo {pages} {768--779} (\bibinfo {year} {2014})}\BibitemShut {NoStop}%
\bibitem [{\citenamefont {Late}\ \emph {et~al.}(2012)\citenamefont {Late},
  \citenamefont {Liu}, \citenamefont {Matte}, \citenamefont {Dravid},\ and\
  \citenamefont {Rao}}]{Late2012}%
  \BibitemOpen
  \bibfield  {author} {\bibinfo {author} {\bibfnamefont {D.~J.}\ \bibnamefont
  {Late}}, \bibinfo {author} {\bibfnamefont {B.}~\bibnamefont {Liu}}, \bibinfo
  {author} {\bibfnamefont {H.~S. S.~R.}\ \bibnamefont {Matte}}, \bibinfo
  {author} {\bibfnamefont {V.~P.}\ \bibnamefont {Dravid}}, \ and\ \bibinfo
  {author} {\bibfnamefont {C.~N.~R.}\ \bibnamefont {Rao}},\ }\bibfield  {title}
  {\enquote {\bibinfo {title} {Hysteresis in single-layer $\m{MoS_2}$ field
  effect transistors},}\ }\href@noop {} {\bibfield  {journal} {\bibinfo
  {journal} {ACS Nano}\ }\textbf {\bibinfo {volume} {6}},\ \bibinfo {pages}
  {5635--5641} (\bibinfo {year} {2012})}\BibitemShut {NoStop}%
\bibitem [{\citenamefont {Fuhrer}\ and\ \citenamefont
  {Hone}(2013)}]{correspondence2013Fuhrer}%
  \BibitemOpen
  \bibfield  {author} {\bibinfo {author} {\bibfnamefont {M.~S.}\ \bibnamefont
  {Fuhrer}}\ and\ \bibinfo {author} {\bibfnamefont {J.}~\bibnamefont {Hone}},\
  }\bibfield  {title} {\enquote {\bibinfo {title} {Measurement of mobility in
  dual-gated $\mathrm{MoS_2}$ transistors},}\ }\href@noop {} {\bibfield
  {journal} {\bibinfo  {journal} {Nature Nanotechnology}\ }\textbf {\bibinfo
  {volume} {8}},\ \bibinfo {pages} {146--147} (\bibinfo {year}
  {2013})}\BibitemShut {NoStop}%
\bibitem [{\citenamefont {Radisavljevic}\ and\ \citenamefont
  {Kis}(2013{\natexlab{a}})}]{correspondence2013Radisavljevic}%
  \BibitemOpen
  \bibfield  {author} {\bibinfo {author} {\bibfnamefont {B.}~\bibnamefont
  {Radisavljevic}}\ and\ \bibinfo {author} {\bibfnamefont {A.}~\bibnamefont
  {Kis}},\ }\bibfield  {title} {\enquote {\bibinfo {title} {Reply to
  "measurement of mobility in dual-gated $\mathrm{MoS_2}$ transistors"},}\
  }\href@noop {} {\bibfield  {journal} {\bibinfo  {journal} {Nature
  Nanotechnology}\ }\textbf {\bibinfo {volume} {8}},\ \bibinfo {pages}
  {147--148} (\bibinfo {year} {2013}{\natexlab{a}})}\BibitemShut {NoStop}%
\bibitem [{\citenamefont {Radisavljevic}\ and\ \citenamefont
  {Kis}(2013{\natexlab{b}})}]{Radisavljevic2013}%
  \BibitemOpen
  \bibfield  {author} {\bibinfo {author} {\bibfnamefont {B.}~\bibnamefont
  {Radisavljevic}}\ and\ \bibinfo {author} {\bibfnamefont {A.}~\bibnamefont
  {Kis}},\ }\bibfield  {title} {\enquote {\bibinfo {title} {Mobility
  engineering and a metal--insulator transition in monolayer
  $\mathrm{MoS_2}$},}\ }\href@noop {} {\bibfield  {journal} {\bibinfo
  {journal} {Nature Materials}\ }\textbf {\bibinfo {volume} {12}},\ \bibinfo
  {pages} {815--820} (\bibinfo {year} {2013}{\natexlab{b}})}\BibitemShut
  {NoStop}%
\bibitem [{\citenamefont {Nguyen}\ \emph {et~al.}(2015)\citenamefont {Nguyen},
  \citenamefont {Sharma}, \citenamefont {Scott}, \citenamefont {Preciado},
  \citenamefont {Klee}, \citenamefont {Sun}, \citenamefont {Lu}, \citenamefont
  {Barroso}, \citenamefont {Kim}, \citenamefont {Shur}, \citenamefont
  {Akhmatkhanov}, \citenamefont {Gruverman}, \citenamefont {Bartels},\ and\
  \citenamefont {Dowben}}]{Nguyen15}%
  \BibitemOpen
  \bibfield  {author} {\bibinfo {author} {\bibfnamefont {A.}~\bibnamefont
  {Nguyen}}, \bibinfo {author} {\bibfnamefont {P.}~\bibnamefont {Sharma}},
  \bibinfo {author} {\bibfnamefont {T.}~\bibnamefont {Scott}}, \bibinfo
  {author} {\bibfnamefont {E.}~\bibnamefont {Preciado}}, \bibinfo {author}
  {\bibfnamefont {V.}~\bibnamefont {Klee}}, \bibinfo {author} {\bibfnamefont
  {D.}~\bibnamefont {Sun}}, \bibinfo {author} {\bibfnamefont {I.-H.~D.}\
  \bibnamefont {Lu}}, \bibinfo {author} {\bibfnamefont {D.}~\bibnamefont
  {Barroso}}, \bibinfo {author} {\bibfnamefont {S.}~\bibnamefont {Kim}},
  \bibinfo {author} {\bibfnamefont {V.~Y.}\ \bibnamefont {Shur}}, \bibinfo
  {author} {\bibfnamefont {A.~R.}\ \bibnamefont {Akhmatkhanov}}, \bibinfo
  {author} {\bibfnamefont {A.}~\bibnamefont {Gruverman}}, \bibinfo {author}
  {\bibfnamefont {L.}~\bibnamefont {Bartels}}, \ and\ \bibinfo {author}
  {\bibfnamefont {P.~A.}\ \bibnamefont {Dowben}},\ }\bibfield  {title}
  {\enquote {\bibinfo {title} {Toward ferroelectric control of monolayer
  $\m{MoS}_2$},}\ }\href@noop {} {\bibfield  {journal} {\bibinfo  {journal}
  {Nano Letters}\ }\textbf {\bibinfo {volume} {15}},\ \bibinfo {pages}
  {3364--3369} (\bibinfo {year} {2015})}\BibitemShut {NoStop}%
\bibitem [{\citenamefont {Preciado}\ \emph {et~al.}(2015)\citenamefont
  {Preciado}, \citenamefont {Sch{\"{u}}lein}, \citenamefont {Nguyen},
  \citenamefont {Barroso}, \citenamefont {Isarraraz}, \citenamefont {von Son},
  \citenamefont {Lu}, \citenamefont {Michailow}, \citenamefont {M{\"{o}}ller},
  \citenamefont {Klee}, \citenamefont {Mann}, \citenamefont {Wixforth},
  \citenamefont {Bartels},\ and\ \citenamefont {Krenner}}]{Preciado2015}%
  \BibitemOpen
  \bibfield  {author} {\bibinfo {author} {\bibfnamefont {E.}~\bibnamefont
  {Preciado}}, \bibinfo {author} {\bibfnamefont {F.~J.}\ \bibnamefont
  {Sch{\"{u}}lein}}, \bibinfo {author} {\bibfnamefont {A.~E.}\ \bibnamefont
  {Nguyen}}, \bibinfo {author} {\bibfnamefont {D.}~\bibnamefont {Barroso}},
  \bibinfo {author} {\bibfnamefont {M.}~\bibnamefont {Isarraraz}}, \bibinfo
  {author} {\bibfnamefont {G.}~\bibnamefont {von Son}}, \bibinfo {author}
  {\bibfnamefont {I.-H.}\ \bibnamefont {Lu}}, \bibinfo {author} {\bibfnamefont
  {W.}~\bibnamefont {Michailow}}, \bibinfo {author} {\bibfnamefont
  {B.}~\bibnamefont {M{\"{o}}ller}}, \bibinfo {author} {\bibfnamefont
  {V.}~\bibnamefont {Klee}}, \bibinfo {author} {\bibfnamefont {J.}~\bibnamefont
  {Mann}}, \bibinfo {author} {\bibfnamefont {A.}~\bibnamefont {Wixforth}},
  \bibinfo {author} {\bibfnamefont {L.}~\bibnamefont {Bartels}}, \ and\
  \bibinfo {author} {\bibfnamefont {H.~J.}\ \bibnamefont {Krenner}},\
  }\bibfield  {title} {\enquote {\bibinfo {title} {Scalable fabrication of a
  hybrid field-effect and acousto-electric device by direct growth of monolayer
  $\m{MoS_2/LiNbO_3}$},}\ }\href {\doibase 10.1038/ncomms9593} {\bibfield
  {journal} {\bibinfo  {journal} {Nature Communications}\ }\textbf {\bibinfo
  {volume} {6}},\ \bibinfo {pages} {8593} (\bibinfo {year} {2015})}\BibitemShut
  {NoStop}%
\bibitem [{\citenamefont {Illarionov}\ \emph {et~al.}(2016)\citenamefont
  {Illarionov}, \citenamefont {Rzepa}, \citenamefont {Waltl}, \citenamefont
  {Knobloch}, \citenamefont {Grill}, \citenamefont {Furchi}, \citenamefont
  {Mueller},\ and\ \citenamefont {Grasser}}]{Illarionov2016}%
  \BibitemOpen
  \bibfield  {author} {\bibinfo {author} {\bibfnamefont {Y.~Y.}\ \bibnamefont
  {Illarionov}}, \bibinfo {author} {\bibfnamefont {G.}~\bibnamefont {Rzepa}},
  \bibinfo {author} {\bibfnamefont {M.}~\bibnamefont {Waltl}}, \bibinfo
  {author} {\bibfnamefont {T.}~\bibnamefont {Knobloch}}, \bibinfo {author}
  {\bibfnamefont {A.}~\bibnamefont {Grill}}, \bibinfo {author} {\bibfnamefont
  {M.~M.}\ \bibnamefont {Furchi}}, \bibinfo {author} {\bibfnamefont
  {T.}~\bibnamefont {Mueller}}, \ and\ \bibinfo {author} {\bibfnamefont
  {T.}~\bibnamefont {Grasser}},\ }\bibfield  {title} {\enquote {\bibinfo
  {title} {{The role of charge trapping in MoS 2 /SiO 2 and MoS 2 /hBN
  field-effect transistors}},}\ }\href {\doibase 10.1088/2053-1583/3/3/035004}
  {\bibfield  {journal} {\bibinfo  {journal} {2D Materials}\ }\textbf {\bibinfo
  {volume} {3}},\ \bibinfo {pages} {035004} (\bibinfo {year}
  {2016})}\BibitemShut {NoStop}%
\bibitem [{\citenamefont {Shayegan}\ \emph {et~al.}(1988)\citenamefont
  {Shayegan}, \citenamefont {Goldman}, \citenamefont {Jiang}, \citenamefont
  {Sajoto},\ and\ \citenamefont {Santos}}]{Shayegan1988}%
  \BibitemOpen
  \bibfield  {author} {\bibinfo {author} {\bibfnamefont {M.}~\bibnamefont
  {Shayegan}}, \bibinfo {author} {\bibfnamefont {V.~J.}\ \bibnamefont
  {Goldman}}, \bibinfo {author} {\bibfnamefont {C.}~\bibnamefont {Jiang}},
  \bibinfo {author} {\bibfnamefont {T.}~\bibnamefont {Sajoto}}, \ and\ \bibinfo
  {author} {\bibfnamefont {M.}~\bibnamefont {Santos}},\ }\bibfield  {title}
  {\enquote {\bibinfo {title} {Growth of low-density two-dimensional electron
  system with very high mobility by molecular beam epitaxy},}\ }\href {\doibase
  10.1063/1.99219} {\bibfield  {journal} {\bibinfo  {journal} {Applied Physics
  Letters}\ }\textbf {\bibinfo {volume} {52}},\ \bibinfo {pages} {1086}
  (\bibinfo {year} {1988})}\BibitemShut {NoStop}%
\bibitem [{\citenamefont {Pfeiffer}\ \emph {et~al.}(1989)\citenamefont
  {Pfeiffer}, \citenamefont {West}, \citenamefont {Stormer},\ and\
  \citenamefont {Baldwin}}]{Pfeiffer1989}%
  \BibitemOpen
  \bibfield  {author} {\bibinfo {author} {\bibfnamefont {L.}~\bibnamefont
  {Pfeiffer}}, \bibinfo {author} {\bibfnamefont {K.~W.}\ \bibnamefont {West}},
  \bibinfo {author} {\bibfnamefont {H.~L.}\ \bibnamefont {Stormer}}, \ and\
  \bibinfo {author} {\bibfnamefont {K.~W.}\ \bibnamefont {Baldwin}},\
  }\bibfield  {title} {\enquote {\bibinfo {title} {{Electron mobilities
  exceeding $10^7\,\m{cm}^2/\m{Vs}$ in modulation-doped GaAs}},}\ }\href
  {\doibase 10.1063/1.102162} {\bibfield  {journal} {\bibinfo  {journal}
  {Applied Physics Letters}\ }\textbf {\bibinfo {volume} {55}},\ \bibinfo
  {pages} {1888} (\bibinfo {year} {1989})}\BibitemShut {NoStop}%
\end{thebibliography}
\end{document}